\documentclass[]{jfm}

\usepackage{graphicx}
\usepackage{dcolumn}
\usepackage{bm}
\usepackage[utf8]{inputenc}
\usepackage{subfig}
\usepackage{graphicx}
\usepackage{ragged2e}
\usepackage{amsmath}
\usepackage{mathtools}
\usepackage{caption}
\usepackage{color}
\usepackage[most]{tcolorbox}
\usepackage{xcolor}
\usepackage{float}
\usepackage{bigints}
\usepackage[export]{adjustbox}
\usepackage{tabularx}
\usepackage{epstopdf, epsfig}

\usepackage[toc,page]{appendix}
\usepackage{nomencl}
\usepackage{mathtools,amssymb,lipsum}
\usepackage{newtxtext}
\usepackage{newtxmath}
\usepackage{natbib}
\usepackage{hyperref}
\hypersetup{
    colorlinks = true,
    urlcolor   = blue,
    citecolor  = black,
}

\newcommand{\RomanNumeralCaps}[1]
\linenumbers

\title{Dimple, jets and self-similarity in nonlinear capillary waves}

\author{Lohit Kayal\aff{1}
  ,
  Saswata Basak\aff{1}
 \and Ratul Dasgupta\aff{1}\corresp{\email{dasgupta.ratul@gmail.com}}}

\affiliation{\aff{1}Chemical Engineering, Indian Institute of Technology, Bombay, India}

\begin{document}
\maketitle
\newcommand{\mj}{{\mathrm{J}}}
\begin{abstract}
Numerical studies of dimple and jet formation from a collapsing cavity often model the initial cavity shape as a truncated sphere, mimicking a bursting bubble. In this study, we present a minimal model containing only nonlinear inertial and capillary forces, which produces dimples and jets from a collapsing, capillary wave trough. The trough develops from an initial perturbation, chosen to be an eigen-mode to the linearised $\mathcal{O}(\epsilon)$ problem ($\epsilon$ is the nondimensional amplitude). We explain the physical mechanism of dimple formation and demonstrate that, for moderate $\epsilon$ the sharp dimple seen in simulations is well captured by the weakly nonlinear $\mathcal{O}(\epsilon^3)$ theory developed here. For $\epsilon >> 1$ the regime is strongly nonlinear spreading surface energy into many modes and the precursor dimple now develops into a sharply rising jet. Here, simulations reveal a novel localised window (in space and time) where the jet evolves self-similarly following inviscid \cite{keller1983surface} scales. We develop an analogy of this regime to a self-similar solution of the first kind, for linearised, capillary waves. Our first principles study demonstrates that at sufficiently small scales, dimples and jets form due to radial focussing of capillary waves requiring (nonlinear) inertial and capillary contributions, sans viscous or gravitational interventions. 
\end{abstract}

\begin{keywords}
\end{keywords}

\section{Introduction}
\label{sec:intro}

\begin{figure}
	\centering
	\includegraphics[scale=0.1]{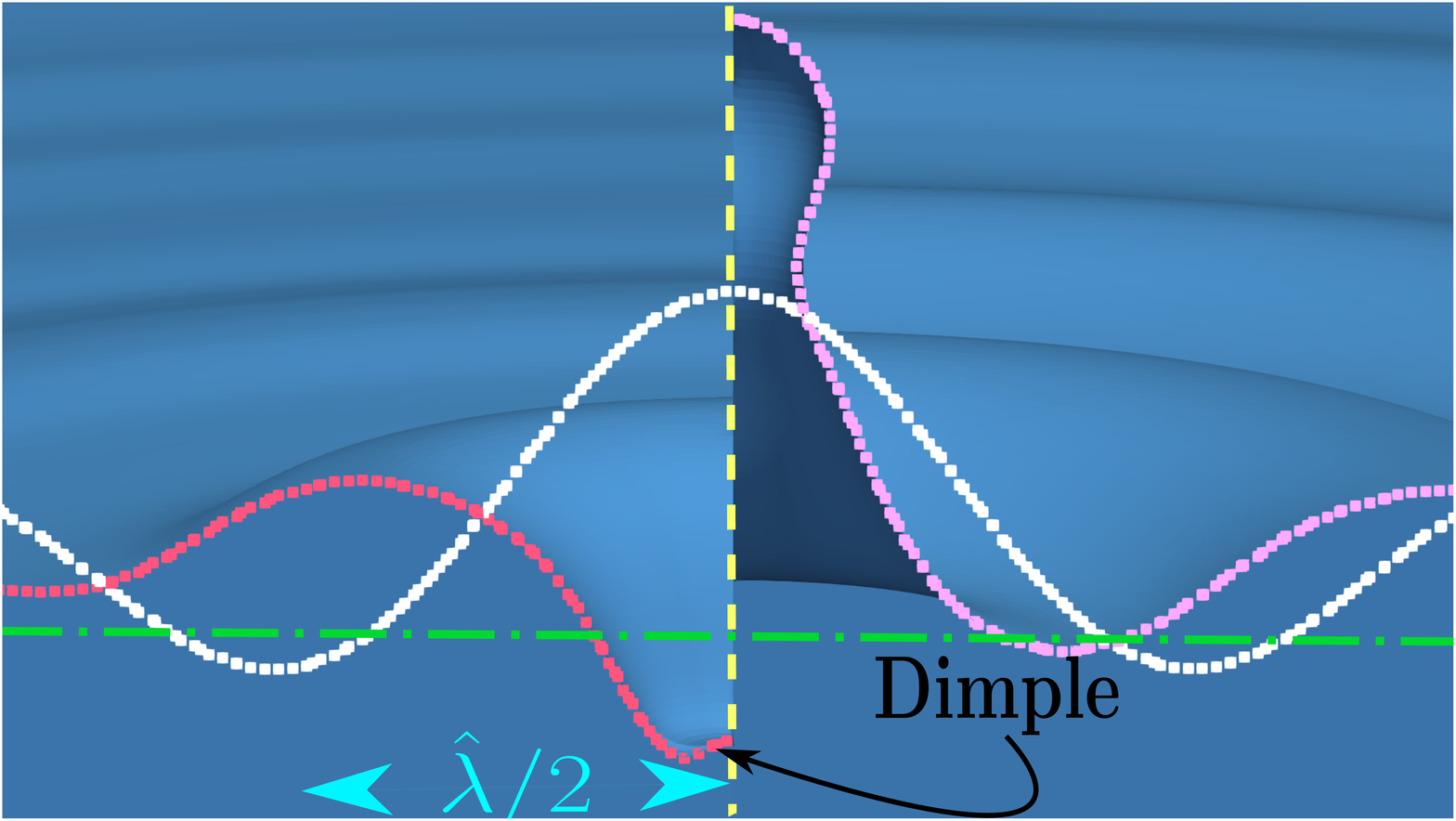}
	\caption{(Vertical dashed yellow line) symmetry axis, (dot-dashed green line) Unperturbed interface, (white dotted curve) initial interface perturbation $\hat{\eta}(\hat{r},0) = \hat{a}_0\mj_{0}\left(l_q\dfrac{\hat{r}}{\hat{R}}\right)$ measured from the unperturbed level, (red dotted curve) trough with precursor dimple at $\hat{t}=0.7947\,\hat{T}$, (pink dotted curve) Rising jet at $\hat{t}=1.0526\,\hat{T}$. Here $\epsilon \approx 2.3,\; l_{15}=47.9$, the full simulation domain is not shown and $\hat{T}$ is the linear time period.} 
	\label{fig1}
\end{figure}

Nonlinear surface waves ranging in wavelengths from metres to microns, remain a bountiful source of curious observations. A phenomena of sustained attention has been jet formation during cavity collapse \citep{kientzler1954photographic}. Due to its relevance to applications such as generation of sea-spray from bursting oceanic bubbles \citep{blanchard1963electrification} or the spread of aroma \citep{seon2017effervescence}, research on this has been sustained \citep{blanchard1963electrification,macintyre1972flow} and intense (\citet{zeff2000singularity,duchemin2002jet,bolanos2008axisymmetric,gordillo2008axisymmetric,ganan2017revision,lai2018bubble,gordillo2019capillary}) continuing actively into the present (\citet{ji2021compound}). The event appears quite ubiquitously and we refer the reader to table $1$ in \cite{basak2021jetting} where a comprehensive literature has been summarised. An estimate of the span of lengths where these jets are observed is instructive. Consider the `spike wave' rising upto $6$ m in height (figure $1$ in \cite{mcallister2022wave}) at the centre of a circular wave tank. At these large scales, the jet dynamics is governed by fluid inertia and gravity \citep{mcallister2022wave}. At the lower extreme in length scales, surface tension dictates the dynamics of an ejecting ethanol jet rising to about $100$ microns due to the collapse of a $35$ micron bubble (figure $1$, top row in \citet{lee2011size} and supplementary movie $1$). In this study, we focus on small-scale jets where the dominant restoring force is surface tension with gravity playing a negligible role. In contrast to prior numerical studies, which obtain the jet by deforming the initial interface in the shape of a truncated spherical cavity \citep{duchemin2002jet,lai2018bubble}, the initial deformation here is chosen to be an eigen-mode to the linearised problem viz. the zeroth order Bessel function (fig. \ref{fig1}). This choice of the initial condition is crucial from a theoretical perspective as we justify below. 

From a modal expansion view-point, a truncated cavity with relatively sharp corners and overhang (fig. 3, lower panel in \cite{duchemin2002jet}) while being a faithful representation of an incipient bursting bubble, is a multi-valued, highly nonlinear surface deformation which does not lend itself to eigen-function expansion (in the radial coordinate). In contrast, the smooth perturbation that we study here excites a single eigen-mode initially. It will be seen that similar to the truncated cavity \citep{duchemin2002jet}, this initial perturbation produces a dimple at moderate amplitude giving way to a jet at larger amplitudes. The initial condition further justifies its choice by yielding to a first-principles (weakly) nonlinear analysis employing eigen-function expansion. The advantage of the resultant theory, devoid of any fitting parameters, is that it describes the onset of dimple formation from a capillary wave trough. In addition, it also provides a simple physical picture explaining the birth of the dimple due to wave focussing. In previous studies from our group, this initial condition has been explored in two regimes viz. the \textit{inviscid, gravity-capillarity} regime in \cite{basak2021jetting} where an $\mathcal{O}(\epsilon^2)$ theory was presented and the \textit{viscous, linear, gravity-capillarity regime} in \cite{farsoiya2017axisymmetric} where the $\mathcal{O}(\epsilon)$ linear problem was studied. In particular, it was shown in \cite{basak2021jetting} that the while the $\mathcal{O}(\epsilon^2)$ theory can describe the jet qualitatively, it does not capture the onset of sharp dimple formation seen in our numerical simulations. An additional difficulty with the capillarity-gravity calculation in \cite{basak2021jetting} is the presence of singularities in the theory due to second harmonic resonance. 

Our current nonlinear minimal model based on capillary and inertial forces only, surmounts the deficiency of the second order theory of \cite{basak2021jetting}, obviates its singularities \citep{kochurin2020numerical} and allows mechanistic interpretation of dimple formation. We demonstrate from first principles that a minimal description of dimple and jet formation at sufficiently small scales, needs to retain only nonlinear curvature and inertial contributions, and notably no contributions from gravitational or viscous forces are necessary. The absence of viscosity is particularly important, as in recent literature there has been significant debate on the physical mechanism of jet formation and the role of viscosity in bubble bursting \citep{ganan2017revision,PhysRevLett.121.269401,ganan2018comment,gordillo2019capillary,ganan2021physics}. While it is known now that the fastest jets from bubble bursting occur at finite viscosity \citep{duchemin2002jet,deike2018dynamics}, our study demonstrates that nonlinear inertial and capillary forces are \textit{necessary and sufficient} ingredients for the generation of these dimples and jets. The nonlinear capillary waves and the accompanying flow generated in our simulations, arise purely due to initial interfacial curvature requiring accurate estimation of the same via the nonlinear theory. In what follows, gravity effects are neglected by restricting attention to wavelengths much smaller than the air-water capillary length scale ($\approx 2.7$ mm). Similarly, viscous effects are ignored as the capillary-viscous length scale for air-water $\rho\nu^2/T$ $\approx 0.01$ microns, \citep{duchemin2002jet} is much smaller than wavelengths of interest.

Fig. \ref{fig1} is a superposition of four snapshots (at different times) obtained from axisymmetric simulations of the inviscid, incompressible Euler's equation with surface tension using the open-source code Basilisk \citep{popinet2014basilisk}. The undisturbed interface is flat (dot-dashed green line) with quiescent water in a cylindrical container of radius $\hat{R}_0$.  The dotted white curve is the initial interface distortion $\hat{\eta}(\hat{r},0) = \hat{a}_0\mj_{0}\left(l_q\dfrac{\hat{r}}{\hat{R}_0}\right)$, $\mj_0(\cdot)$ being the zeroth order Bessel function and $q \in \mathbb{Z}^{+}$ indicates number of zero crossings of $\mj_0(\cdot)$ within the computational domain. The blue surface in the background is the axisymmetric rendition of the interface, at the instant of dimple formation (left half) and for the rising jet (right half). Evolving from the initial perturbation (white dotted curve), the interface develops a trough and as this trough rises, a dimple forms at its base (red dotted curve). This dimple subsequently develops into a jet (pink dotted curve), rising signifcantly more than the initial perturbation at $\hat{r}=0$. For simplicity, the container is assumed to be deep, implying that the longest capillary wave ($\hat{\lambda}_{\text{max}}$) in our domain satisfies $\hat{\lambda}_{\text{max}} << \hat{H}$, where $\hat{H}$ is the undisturbed water depth (dimensional variables have a hat). The simulations are described by two non-dimensional parameters viz. $\epsilon \equiv \hat{a}_0\frac{l_q}{\hat{R}_0}$ and a number $l_q$ with $q$ being the index of the primary Bessel mode that we excite initially. These provide a non-dimensional measure of the initial perturbation amplitude $\hat{a}_0$ and wavelength $\hat{\lambda}$ (see fig. \ref{fig1}) respectively. The numbers $l_q$ satisfy $\mj_1(l_q)=0$ \citep{weisstein2020bessel}, necessary for respecting the no-penetration condition at $\hat{r}=\hat{R}_0$ (see next section). A rough estimate of the width of the crest in fig. \ref{fig1} viz. $\hat{\lambda} \approx \dfrac{2\pi \hat{R}_0}{l_q}$, is obtained from the asymptotic expression of $\mj_{0}(\cdot)$ at large radius \citep{abramowitz1972handbook}. We may treat $\hat{\lambda}$ as a measure of the initial perturbation wavelength, noting that this is an approximation as the zeros of $\mj_{0}(\cdot)$ are not equally spaced. In order to minimise container wall effects on dimple and jet formation, we require $\frac{\hat{R}_0}{\hat{\lambda}} = \frac{l_q}{2\pi} >> 1$ which can be ensured by choosing $q$ to be sufficiently large. In this study, we have chosen $q \geq 15$ to stay consistent with this requirement.
\begin{figure}
\centering
	\subfloat[]{\includegraphics[scale=0.1]{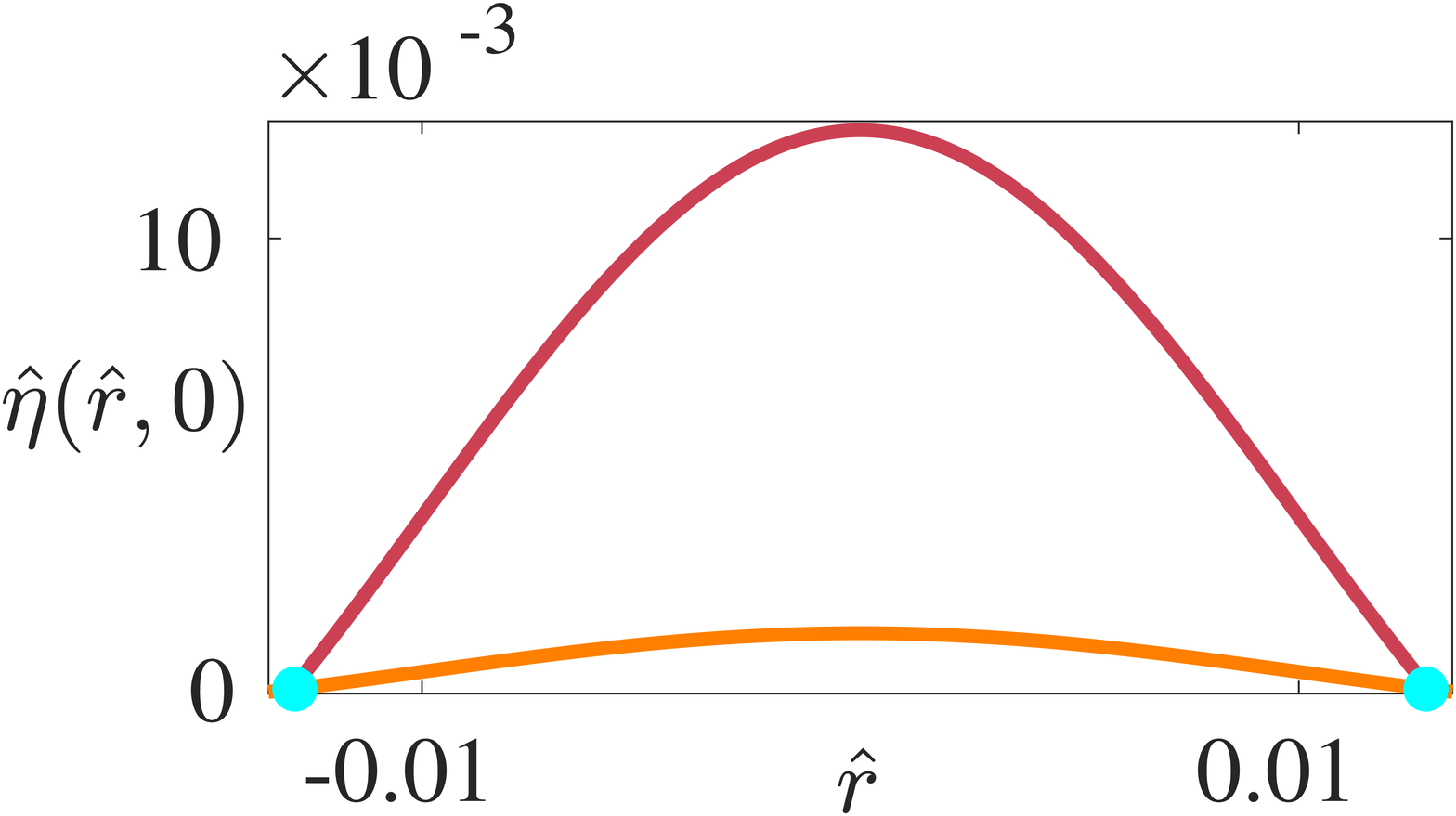}\label{fig2a}}
	\subfloat[]{\includegraphics[scale=0.09]{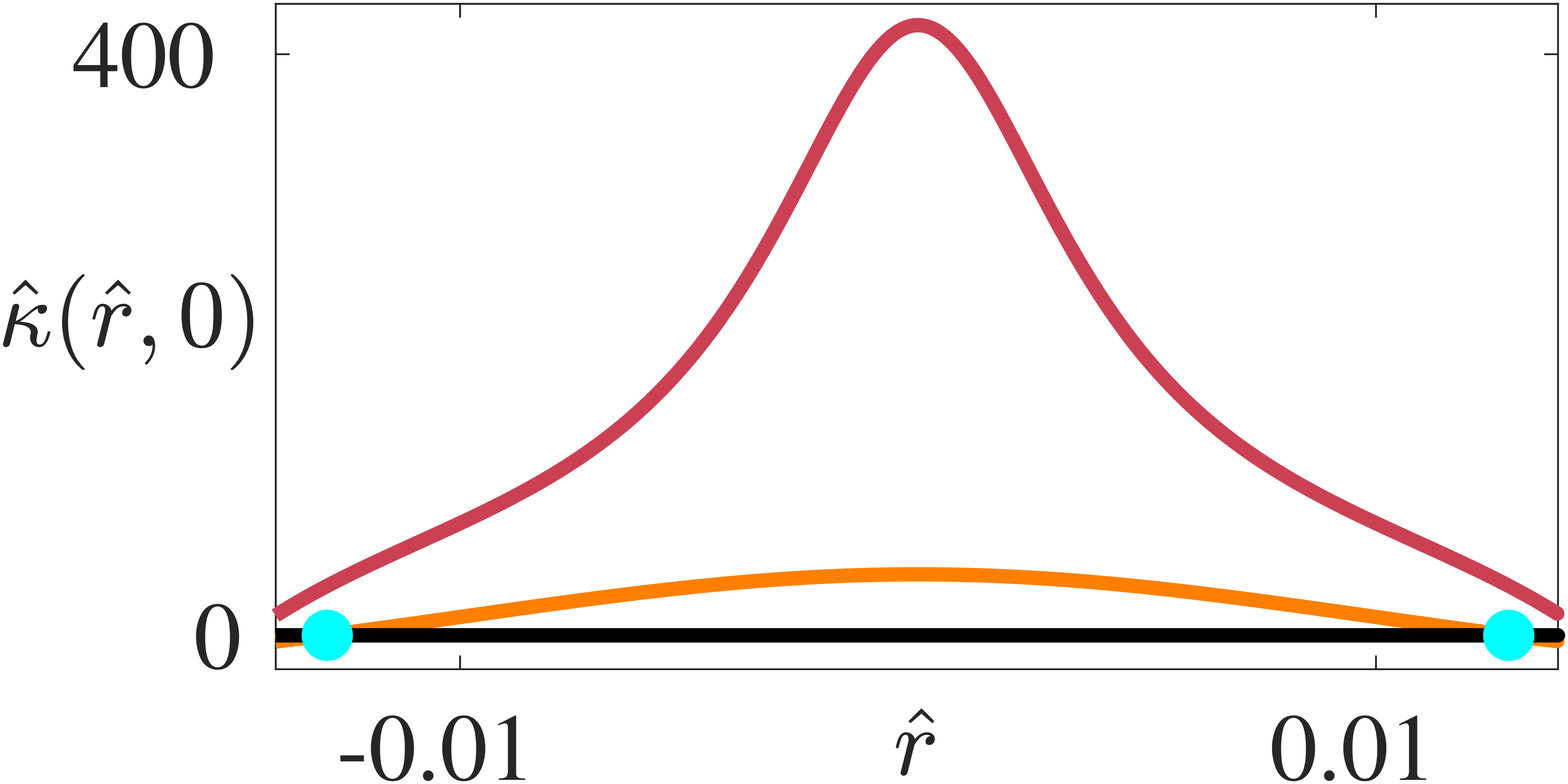}\label{fig2b}}\\
	\subfloat[]{\includegraphics[scale=0.12]{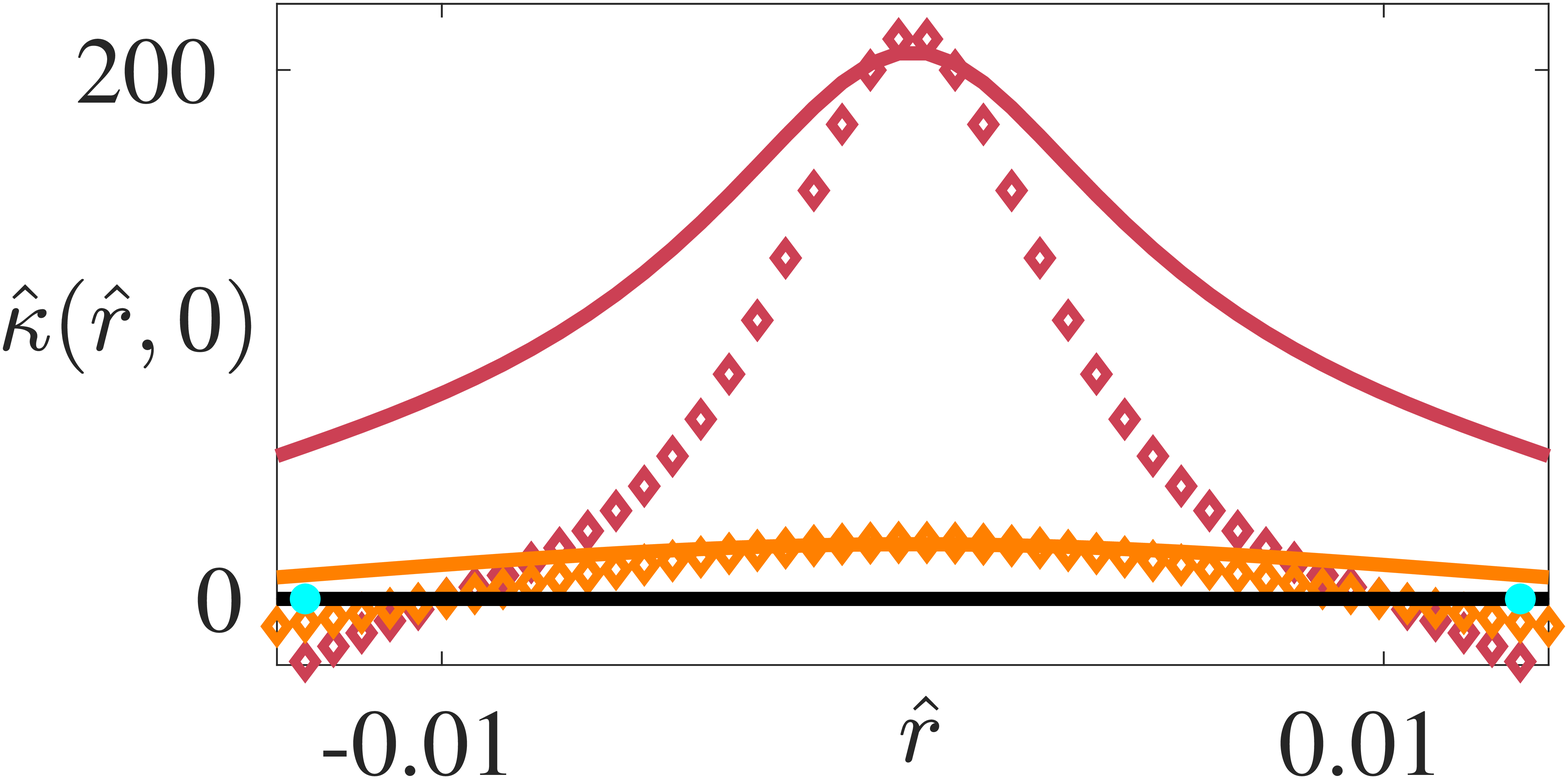}\label{fig2c}}
	\subfloat[]{\includegraphics[scale=0.12]{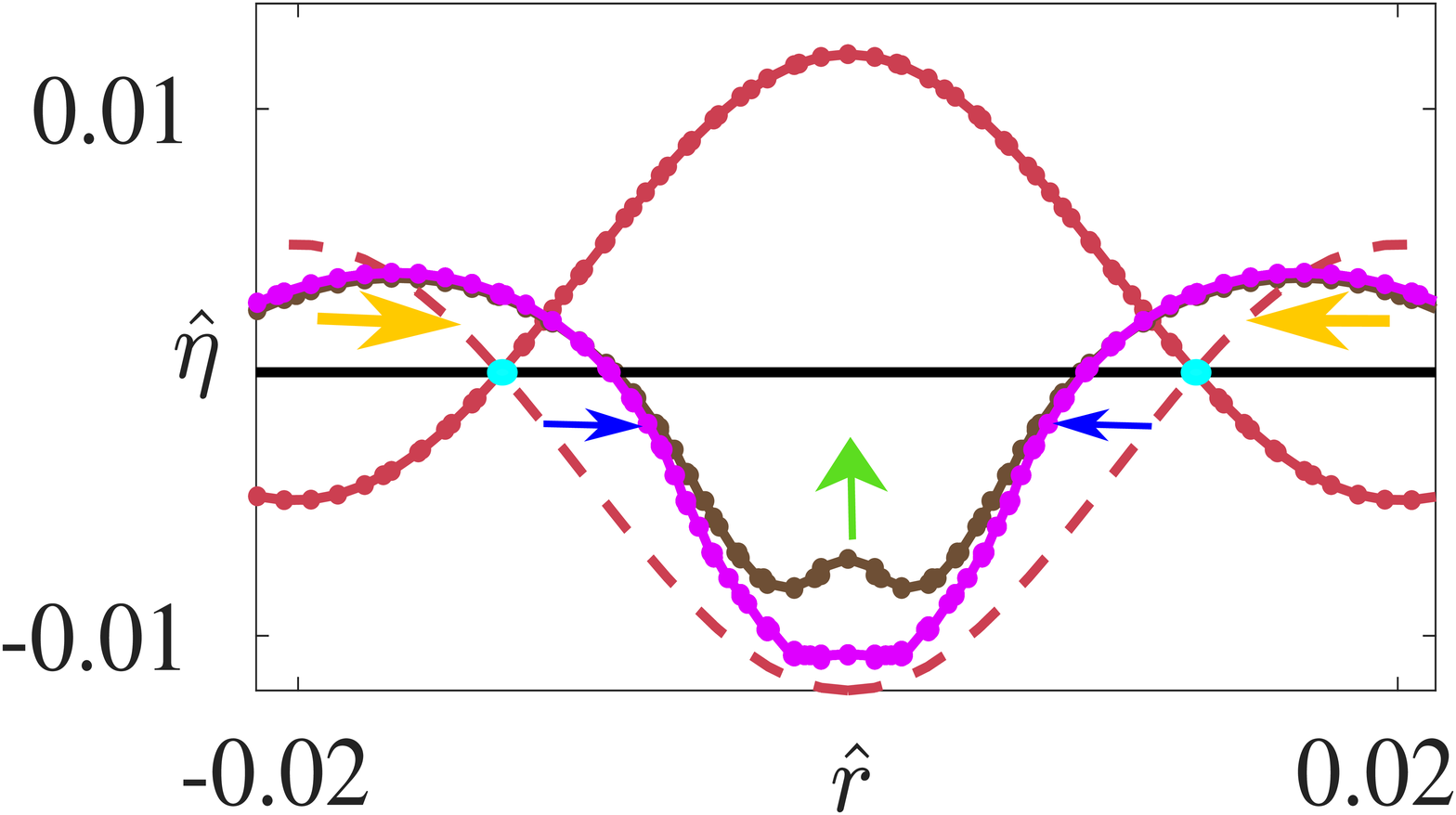}\label{fig2d}}
	\caption{a) $\eta(\hat{r},0) = \hat{a}_0\mj_{0}\left(l_q\frac{\hat{r}}{\hat{R}_0}\right)$ with (brown curve) $\epsilon\approx 2.3$ and (orange curve) $\epsilon\approx 0.23$. b) Initial curvature $\hat{\kappa}(\hat{r},0)$ with the same color legend as fig. \ref{fig2a}. c) (Solid curves) Radial curvature and (hollow diamonds) axial curvature, same color legend as fig. \ref{fig2a} d) (Brown dotted curve) $\hat{\eta}(\hat{r},0)$, (Purple dotted curve) Flat trough formation, (Deep brown dotted curve) Trough with incipient sharp dimple, (Dashed brown curve) $-\hat{\eta}(\hat{r},0)$ }
	\label{fig2}	
\end{figure}

Fig. \ref{fig2} explains the physical mechanism of dimple formation. Plotted in fig. \ref{fig2a} is $\hat{a}_0\mj_{0}(l_q\frac{\hat{r}}{\hat{R}_0})$ for large and small amplitudes viz. $\epsilon = \hat{a}_0\frac{l_q}{\hat{R}_0} \approx 2.3$ (brown curve) and $\epsilon \approx0.23$ (orange curve) with $l_{24}=74.18$ for both cases. Fig. \ref{fig2b} shows the initial curvature profile $\hat{\kappa}(\hat{r},0)$ for the two Bessel modes in fig. \ref{fig2a}. Fig. \ref{fig2c} decomposes the curvature in fig. \ref{fig2b} into axial (diamonds) and radial curvature contributions (solid lines). With $\hat{\eta}(\hat{r},0) = \hat{a}_0\mj_{0}(l_q\frac{\hat{r}}{\hat{R}_0})$ and $\epsilon << 1$, the linearised approximation to curvature is $-\hat{\eta}_{\hat{r}\hat{r}}(\hat{r},0) - \hat{r}^{-1}\hat{\eta}_{\hat{r}}(\hat{r},0)\propto \hat{a}_0\mj_{0}(l_q\frac{\hat{r}}{\hat{R}_0})$ (subscript $\hat{r}$ indicates partial derivative). Thus for a small amplitude Bessel mode, its curvature profile is also a Bessel mode implying that the curvature is zero at the zero crossings of the mode. Consequently, if the interface is initialised as a small amplitude Bessel mode, we expect a standing wave whose shape at all subsequent time remains proportional to this mode. Note that at small amplitude the curvature being nearly zero at the first zero crossing ($\hat{r} \approx \pm 0.013$, cyan dots in fig. \ref{fig2}) in fig. \ref{fig2a} is an outcome of the near cancellation of axial and radial curvature terms in the exact expression for curvature (fig. \ref{fig2c}, orange curve and hollow orange diamonds). The aforementioned description changes qualitatively as $\epsilon$ approaches unity and exceeds it. For $\epsilon > 1$ ($\epsilon=2.3$), the curvature profile indicated by the brown curve in fig. \ref{fig2b}, appears quite distinct from the interface itself in fig. \ref{fig2a}. Notice that the cancellation between the curvature components at the zero crossing now becomes imperfect (fig. \ref{fig2c}) producing a net positive curvature at the zero crossings. A finite curvature at the zero crossing implies a local interface velocity, which moves the zero crossings radially inwards (see fig. \ref{fig2d}, blue arrows). This radially inward motion results in a trough (purple dotted curve in fig. \ref{fig2d}) which is significantly narrower than what would result, had the initial perturbation evolved as a linearised wave (dashed brown curve in fig. \ref{fig2d}). Fig. \ref{fig2d} also shows that the inward motion of zero crossings, coincides with the inward motion of the two capillary humps (purple dotted curve) indicated by yellow arrows, interpreted here as capillary wave focussing \citep{lai2018bubble}. This focussing produce a region of large curvature at the symmetry axis which triggers the dimple. Analogous focussing of capillary waves is also implicated in the formation of dimples and jets from a bursting bubble \citep{duchemin2002jet,ganan2017revision,PhysRevLett.121.269401}. In the following section, we develop a weakly nonlinear approach to understand the formation of this dimple. The motivation for this lies in the observation that the dimple is already seen at moderate values of $\epsilon \approx 0.8$ (fig. \ref{fig3}), rendering its formation accessible to first principles theory. 
\subsection{\textbf{Weakly nonlinear regime:} $\epsilon  < 1$}
We use $\epsilon \equiv \frac{\hat{a}_0l_q}{\hat{R}_0} < 1$, to solve the potential flow equations with cylindrical, axisymmetric surface tension terms \citep{mathur2007gravity} perturbatively upto $\mathcal{O}(\epsilon^3)$. Equations \ref{eq1}(a-i) are non-dimensionalised using the capillary scales, $l_q/\hat{R}_0$ and $\left(\frac{T l_q^3}{\rho\hat{R}_0^3}\right)^{1/2}$ respectively, $T$ being surface tension, density $\rho$  and container radius $\hat{R}_0$ while subscripts $r$ and $t$ indicate partial derivatives. Equation \ref{eq1}a is the Laplace equation governing the perturbation velocity potential $\phi$, \ref{eq1}(b,c) are the kinematic boundary condition and the Bernoulli equation at the interface respectively. Equations \ref{eq1}(d,e) are the no-penetration and free-edge boundary condition at the wall respectively ensuring that the contact angle is $\pi/2$. Equation \ref{eq1}(f) ensures that the perturbation conserves volume while \ref{eq1}(g-i) represent initial conditions. These equations contain the dimensionless group $\epsilon$ and parameter $l_q$. The analytical procedure is similar to the capillary-gravity case in \cite{basak2021jetting} with the important difference that we obtain corrections upto cubic order here. The necessity of going upto $\mathcal{O}(\epsilon^3)$ is seen from the exact curvature expressions in eqn. \ref{eq1}c. For $\eta(r,0) \propto \epsilon\mj_{0}(r)$, the first nonlinear contribution from curvature appears at $\mathcal{O}(\epsilon^3)$. 
	\begin{subequations}\label{eq1}
			\begin{align}         	
	&\nabla^2\phi = 0,\tag{\theequation a}\\
	&\frac{\partial \eta}{\partial t} + \frac{\partial \eta}{\partial r}\left(\frac{\partial\phi}{\partial r}\right)_{z = \eta} = \left(\frac{\partial\phi}{\partial z}\right)_{z = \eta},\tag{\theequation b} \\
	&\left(\frac{\partial\phi}{\partial t}\right)_{z = \eta} + \frac{1}{2} \left[\left(\frac{\partial\phi}{\partial r}\right)^2 + \left(\frac{\partial\phi}{\partial z}\right)^2\right]_{z=\eta}- \left[\frac{\frac{\partial^2\eta}{\partial r^2}}{\big\{1 + \left(\frac{\partial\eta}{\partial r}\right)^2\big\}^{3/2}} + \frac{1}{r}\frac{\frac{\partial\eta}{\partial r}}{\big\{1 + \left(\frac{\partial\eta}{\partial r}\right)^2\big\}^{1/2}}\right] = 0\;\;\text{at}\;\;z = \eta \tag{\theequation c} \\
	&\left(\frac{\partial\phi}{\partial r}\right)_{r=l_q}=\left(\frac{\partial\eta}{\partial r}\right)_{r=l_q} = 0, \;\int_{0}^{l_q} dr r\eta(r,t) = 0,\;\eta(r,0) = \epsilon \mathrm{J_0}(r), \; \frac{\partial\eta}{\partial t}(r,0) =0,\; \phi(r,z,0) = 0 \tag{\theequation d,e,f,g,h,i}
		\end{align}         	
	\end{subequations}
	For eliminating resonant forcing of primary mode ($q$), we have $\tau = t\left[1 + \epsilon^2 \Omega_2 + \ldots\right]$. Following \cite{basak2021jetting}, we expand $\phi(r,z,\tau) = \sum_{i=1}^{3}\epsilon^{i}\phi_i(r,z,\tau)$ and $\eta(r,\tau) = \sum_{i=1}^{3}\epsilon^{i}\eta_i(r,\tau)$. Taylor expanding all $z$ dependent terms about $z$=$0$, we obtain coupled, linear, partial differential equations at every order $\mathcal{O}(\epsilon^{i})$ with non-linearity appearing at $i \geq 2$. Using the Dini series \citep{basak2021jetting} $\phi_i(r,z,\tau) = \sum_{j = 0}^{\infty}p_i^{(j)}(\tau)\exp(\alpha_{j,q} z)\mathrm{J_0}(\alpha_{j,q} r)$ and $\eta_i(r,z,\tau) = \sum_{j = 0}^{\infty}a_i^{(j)}(\tau)\mathrm{J_0}(\alpha_{j,q} r)$ with $\alpha_{j,q} \equiv l_j/l_q$, the task of determining $\phi_i$ and $\eta_i$ becomes that of determining $p_i^{(j)}(\tau)$ and $a_i^{(j)}(\tau)$. These satisfy simple harmonic oscillator equations with in-homogenities at $\mathcal{O}(\epsilon^2)$ and $\mathcal{O}(\epsilon^3)$ which may be solved subject to initial conditions in eqns. \ref{eq1}(g-i). The final expressions for $\eta_1(r,\tau),\eta_2(r,\tau)$ and $\eta_3(r,\tau)$ are provided below (expression for modal coefficients are lengthy and provided in the supplementary material). The expressions for $\eta_2(r,\tau)$ and $\eta_3(r,\tau)$ clearly show the excitation of modes not present initially in the spectrum implying that surface energy injected into a single mode (index $q$) initially, gets re-distributed among other modes and this is crucial for resolving dimple formation. Note the presence of bound and free wave components in expressions for $\eta_{2}(r,\tau),\eta_{3}(r,\tau)$ in eqn. \ref{eq1cont}(b,c) ($\omega_{j,q}^2 \equiv \alpha_{j, q}^3$). 
		\begin{subequations}\label{eq1cont}
		\begin{align}         	
		&\eta_1(r,\tau) = \cos(\tau)\mathrm{J_0}(r),\; \eta_2(r,\tau) = \frac{1}{2}\sum_{j = 1}^{\infty}\left[\zeta_{j,q}^{(1)}\cos(\omega_{j,q}\tau) + \zeta_{j,q}^{(2)}\cos(2\tau) + \zeta_{j,q}^{(3)}\right]\mathrm{J_0}(\alpha_{j,q}r)\nonumber\\
		&\eta_{3}(r, \tau)=\bigg[\mathrm{\mu}^{(q)}\cos (\tau)+\mathrm{\kappa}^{(q)} \cos (3 \tau)\nonumber+ \nonumber \\ 
		&\sum_{m=1, m \neq q}^{\infty}\left\lbrace\mathrm{\gamma}_{m}^{(q)} \cos \left[\left(\omega_{m, q}+1\right) \tau\right]+\mathrm{\chi}_{m}^{(q)} \cos \left[\left(\omega_{m, q}-1\right) \tau\right]\right\rbrace\bigg] \mathrm{J}_{0}(r)\nonumber\\
		&+\sum_{j=1, j \neq q}^{\infty}\bigg[\mathrm{\mu}^{(j)} \cos (\tau)+\mathrm{\kappa}^{(j)} \cos (3 \tau)+\mathrm{\nu}^{(j)} \cos \left(\omega_{j, q} \tau\right)+ \nonumber \\
		&\sum_{m=1, m \neq q}^{\infty}\left\lbrace\mathrm{\gamma}_{m}^{(j)} \cos \left[\left(\omega_{m, q}+1\right) \tau\right]+\mathrm{\chi}_{m}^{(j)} \cos \left[\left(\omega_{m, q}-1\right) \tau\right]\right\rbrace\bigg] \mathrm{J}_{0}\left(\alpha_{j, q} r\right) \tag{\theequation a,b,c}
		\end{align}         	
	\end{subequations}
\begin{figure}
	\centering
	\subfloat[$\hat{t} = 0$]{\includegraphics[scale=0.11]{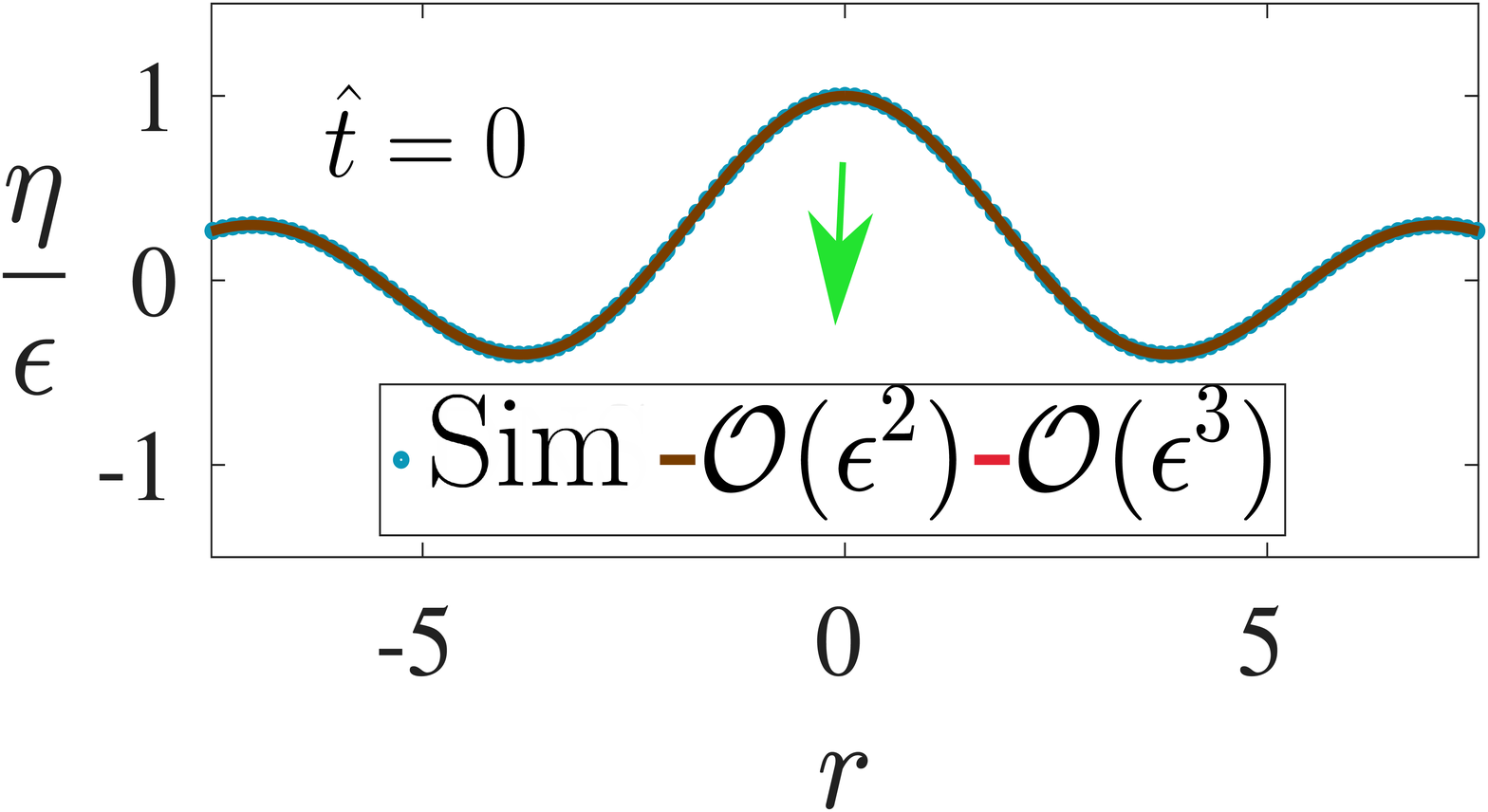}\label{fig3a}}
	\subfloat[$\hat{t} = 0.25\;\hat{T}$]{\includegraphics[scale=0.11]{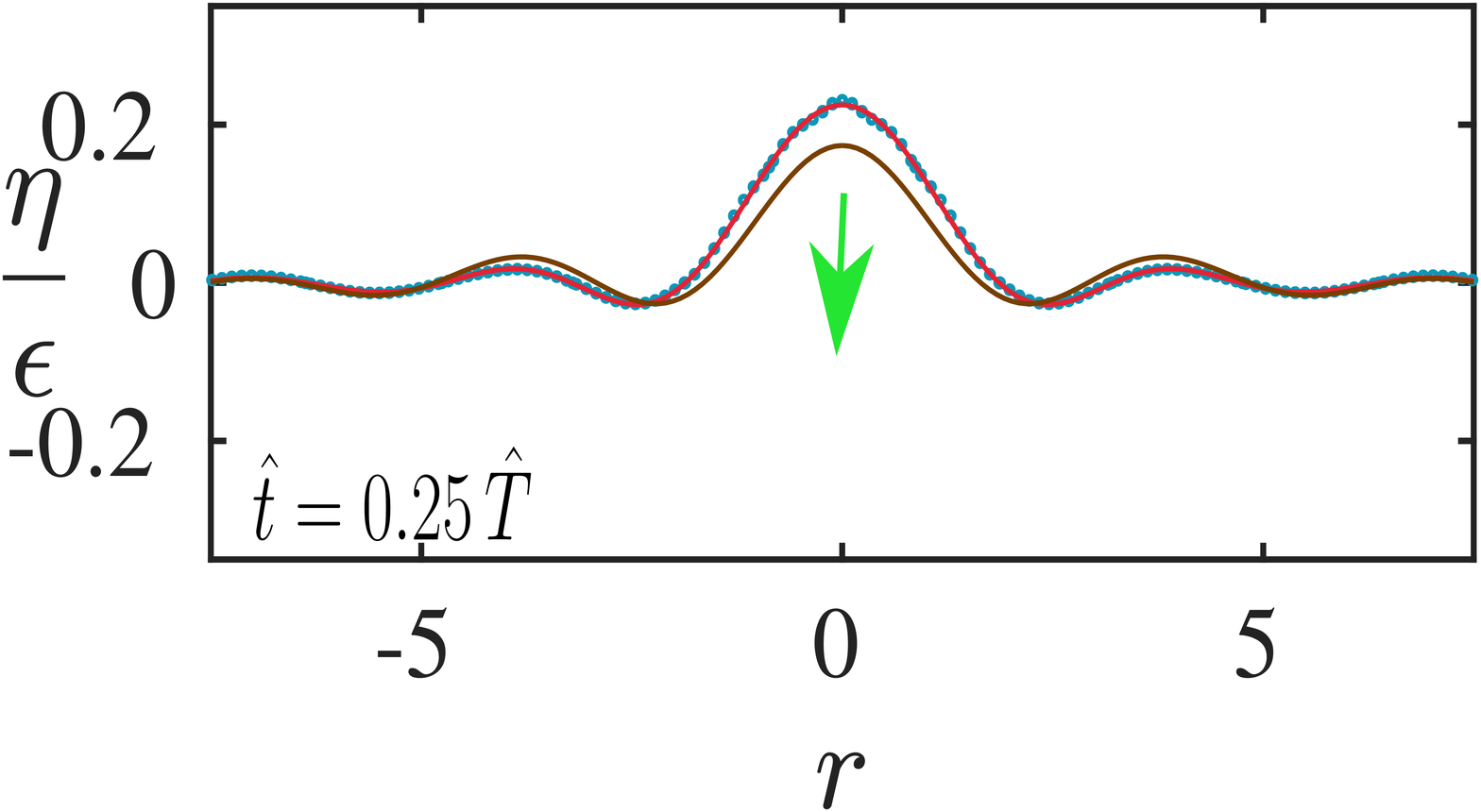}\label{fig3b}}\\
	\subfloat[$\hat{t} = 0.54\;\hat{T}$]{\includegraphics[scale=0.11]{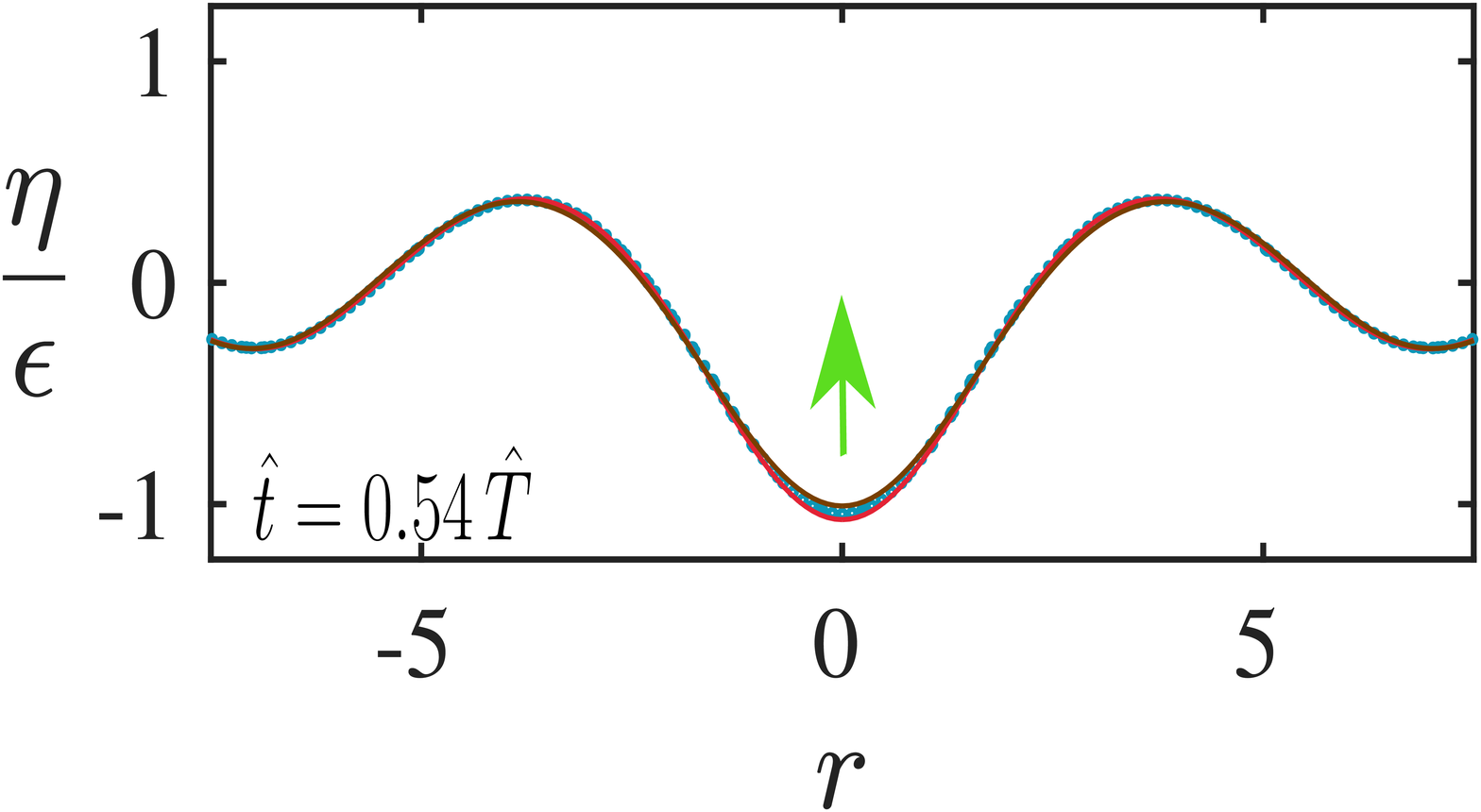}\label{fig3c}}
	\subfloat[$\hat{t} = 0.69\;\hat{T}$]{\includegraphics[scale=0.11]{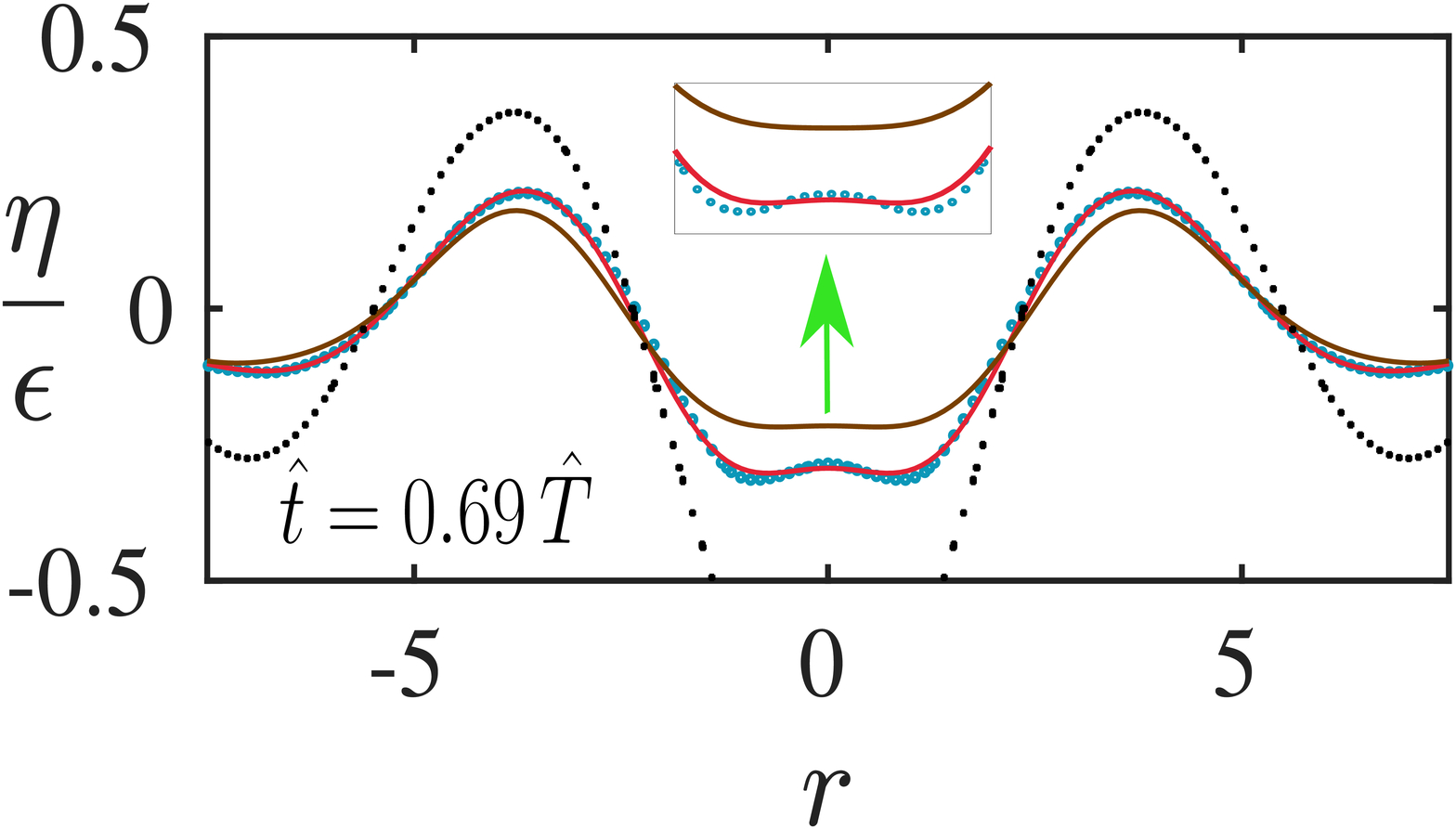}\label{fig3d}}
	\caption{Interface shape at different time instants from simulations and second, third order theory for $\epsilon = 0.8, l_{15} = 47.9$. The green arrows indicate the instantaneous direction of interface motion. Panel (d) (Black dotted curve) Wave trough at an earlier instant before the emergence of dimple.}
	\label{fig3}
\end{figure}

\begin{figure}
	\centering
	\includegraphics[scale=0.1]{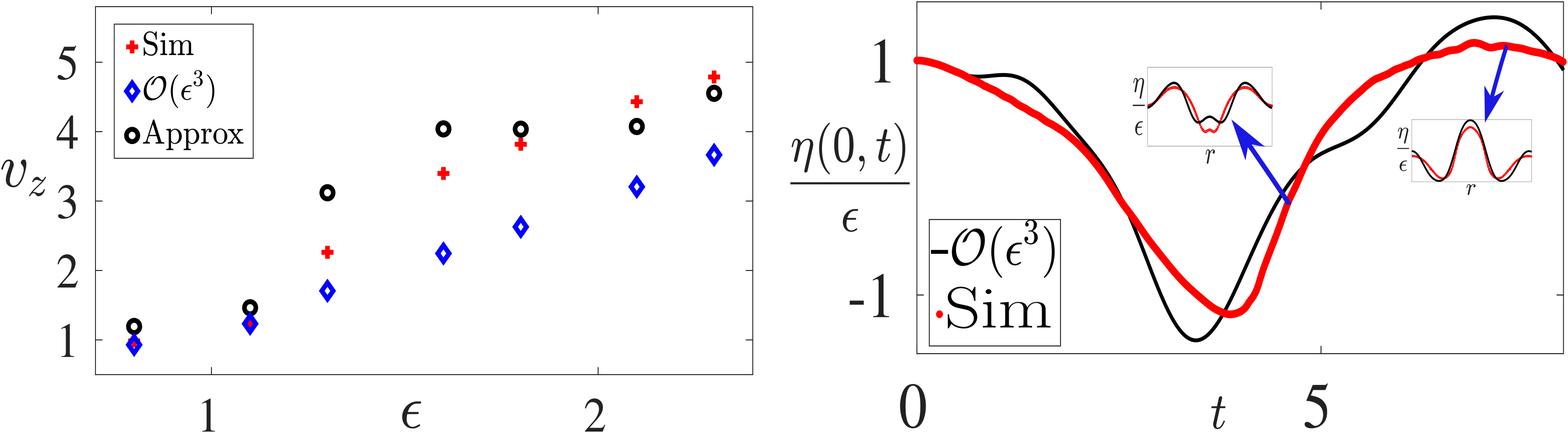}
	\caption{a) Dimple velocity as function of $\epsilon$ for $l_{15}$. (Approx) indicates the two term order of magnitude balance discussed in text. b) Time signal $\epsilon = 1.8, l_{15} = 47.9$. At $\epsilon >> 1$, the weakly nonlinear theory becomes inadequate.}
	\label{fig4}
\end{figure}

Fig. \ref{fig3} compares shape of the interface at different time instants for moderate $\epsilon=0.8$ between theory and simulations. It is clear that the $\mathcal{O}(\epsilon^3)$ theory does significantly better than its second order counterpart, particularly in capturing the instantaneous shape and location of the dimple (fig. \ref{fig3d}). Note that the width of this dimple is substantially smaller than the width of the collapsing wave trough (black dotted curve in fig. \ref{fig3d}). Fig. \ref{fig4} presents the vertical velocity of the dimple at its formation from simulations (red plus) and from $\mathcal{O}(\epsilon^3)$ theory (blue diamonds). Good agreement between the two can be seen upto $\epsilon = 1.1$. Beyond $\epsilon > 1.1$, sharply rising jets are seen in simulations and the weakly nonlinear theory systematically underestimates its velocity. A better estimate may be obtained by a two term order-of-magnitude balance, in eqn. \ref{eq1}c. Approximating the dimple velocity to be predominantly vertical, we have $v_z^2 \sim \frac{T}{\rho}\left(\frac{1}{R_1} + \frac{1}{R_2}\right)$ where $R_1^{-1}$ and $R_2^{-1}$ are estimated from simulations. This is shown in fig. \ref{fig4}, panel (a) (black circles) where it is seen that for $\epsilon >> 1$, this approximation does better than the $\mathcal{O}(\epsilon^3)$ theory. The limitation of the weakly nonlinear approach as $\epsilon >> 1$ is also seen clearly in fig. \ref{fig4} panel (b), where we track the interface in time at $r=0$ for $\epsilon=1.8$. It is seen (insets) that the dimple becomes very narrow with increasing $\epsilon$. Such sharp dimples involves energy transfer to modes with indices $>> 3q$ in the spectrum \cite{basak2021jetting} and falls into the strongly nonlinear regime that we discuss next.
\subsection{\textbf{Strongly nonlinear regime($\epsilon >> 1$): self-similar evolution}}
\begin{figure}	
\centering
	\centering
	\subfloat[]{\includegraphics[scale=0.11]{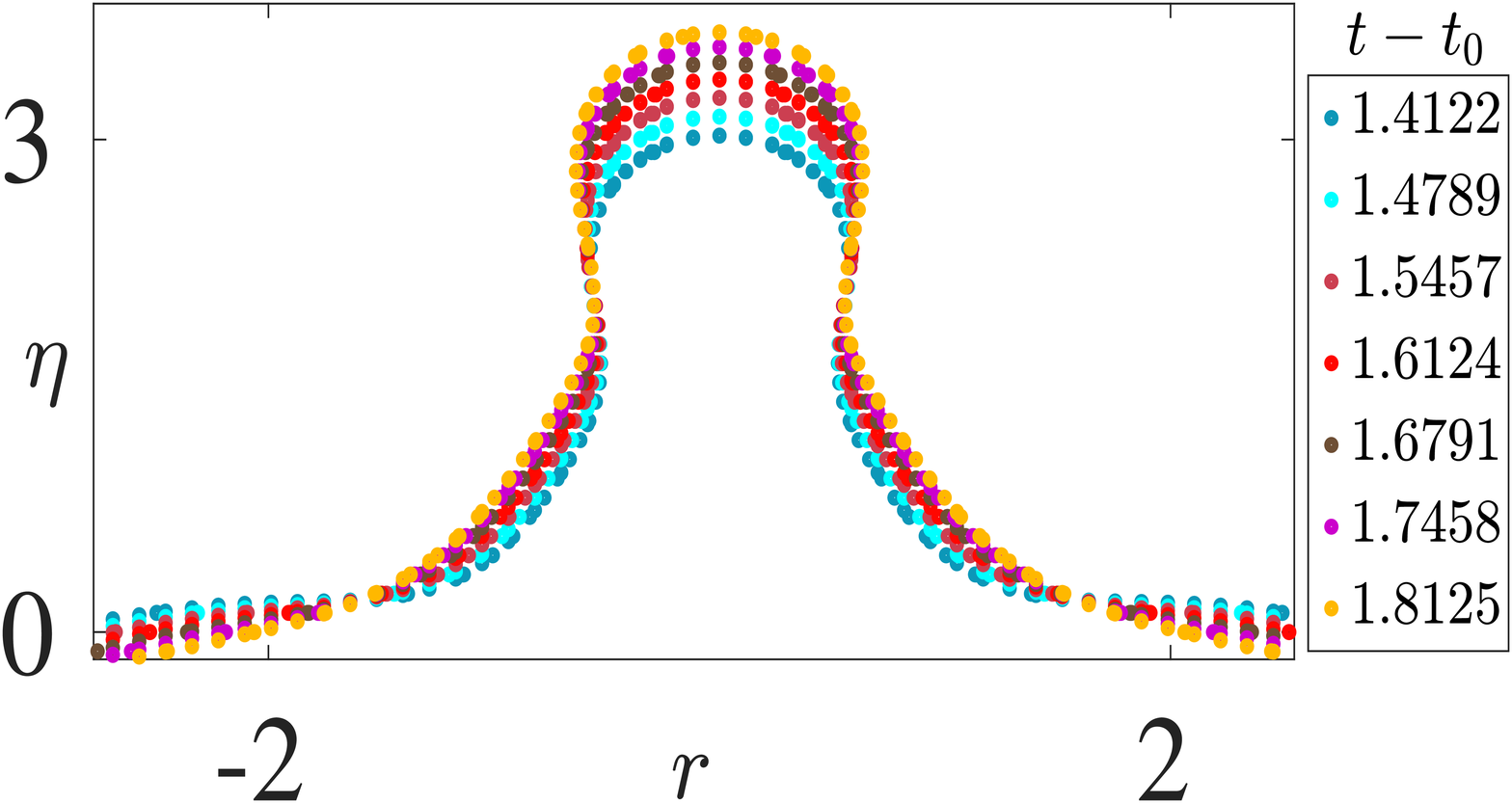}\label{fig5a}}\quad
	\subfloat[]{\includegraphics[scale=0.11]{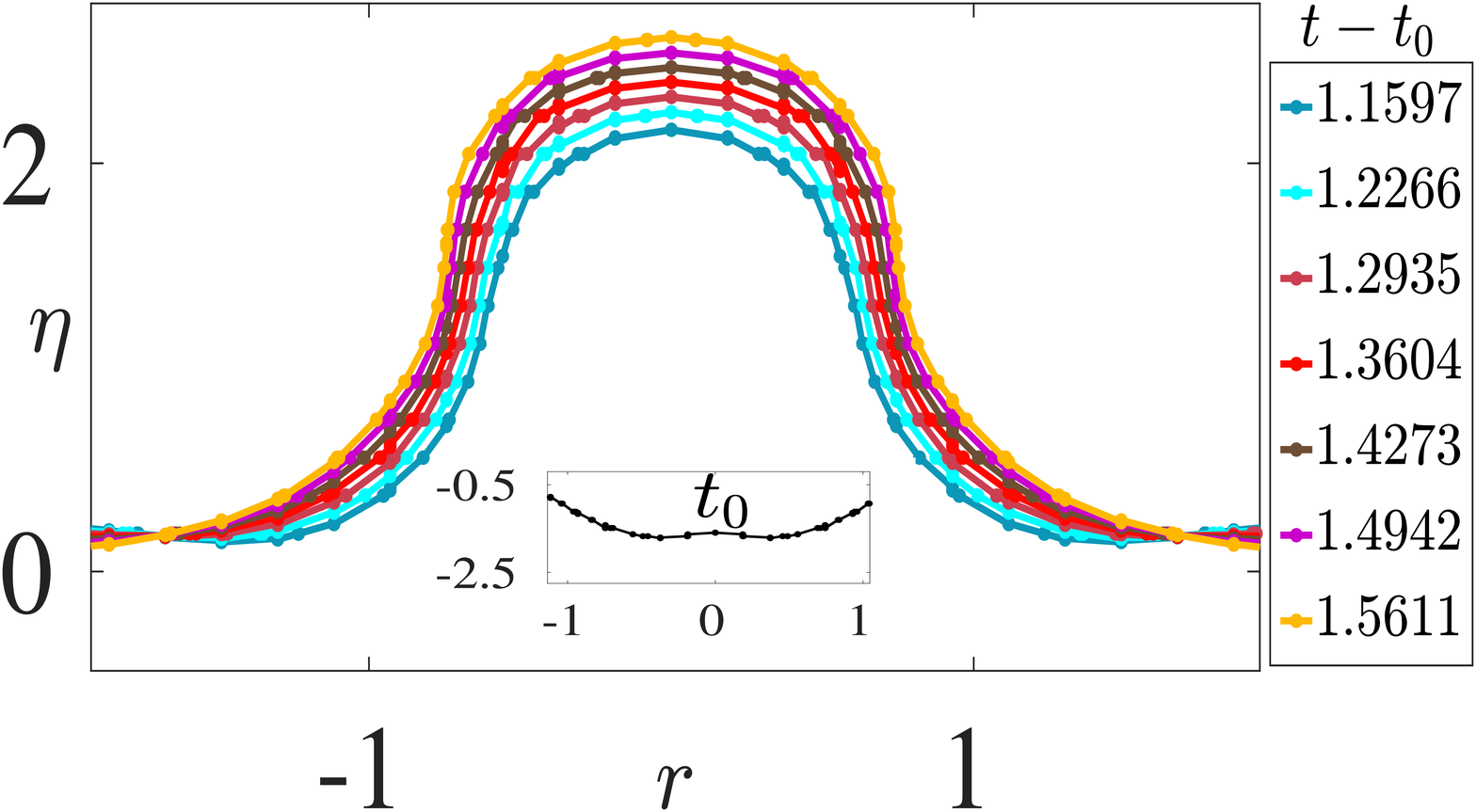}\label{fig5b}}\\
	\subfloat[]{\includegraphics[scale=0.1]{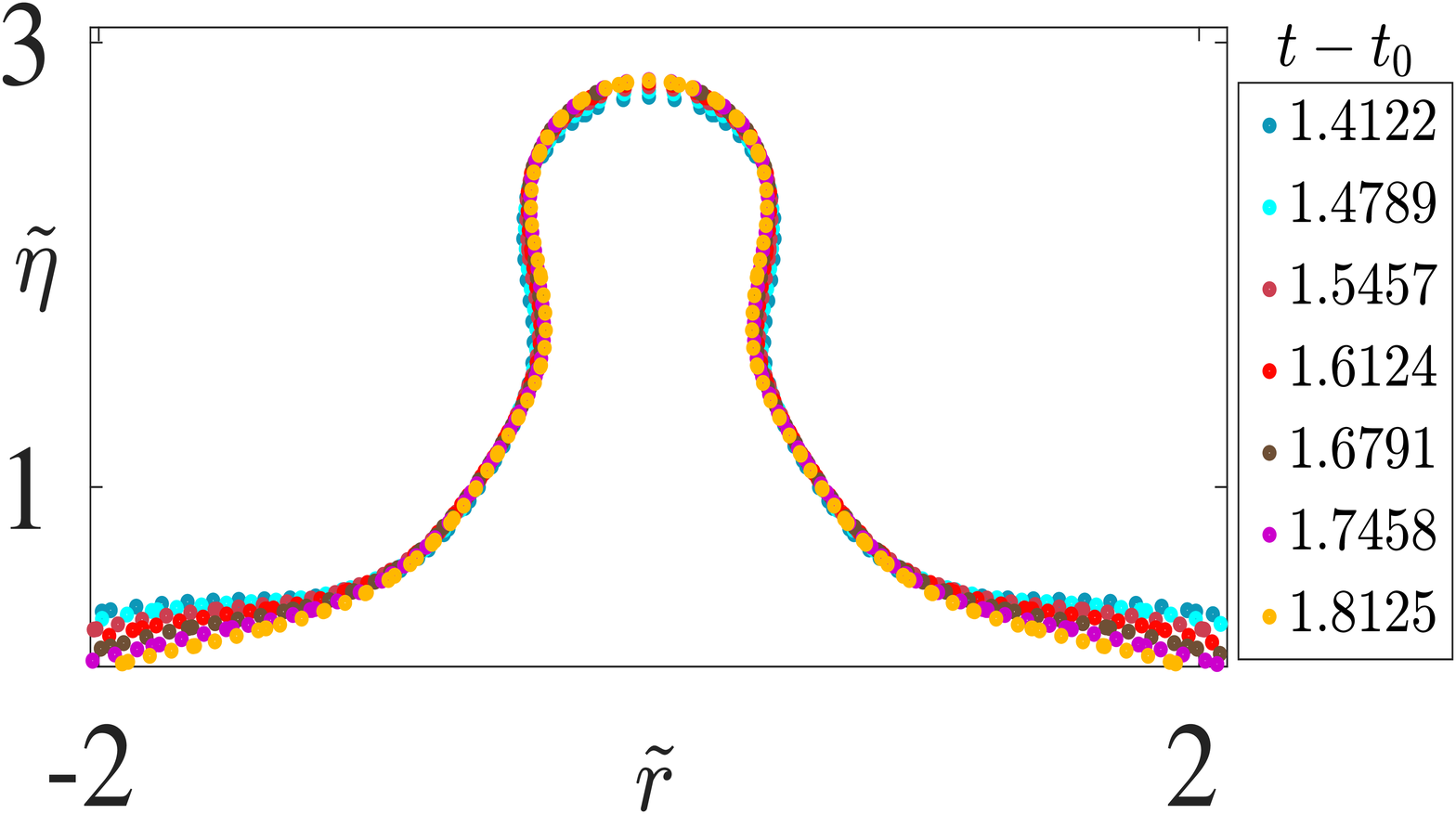}\label{fig5c}}\quad
	\subfloat[]{\includegraphics[scale=0.1]{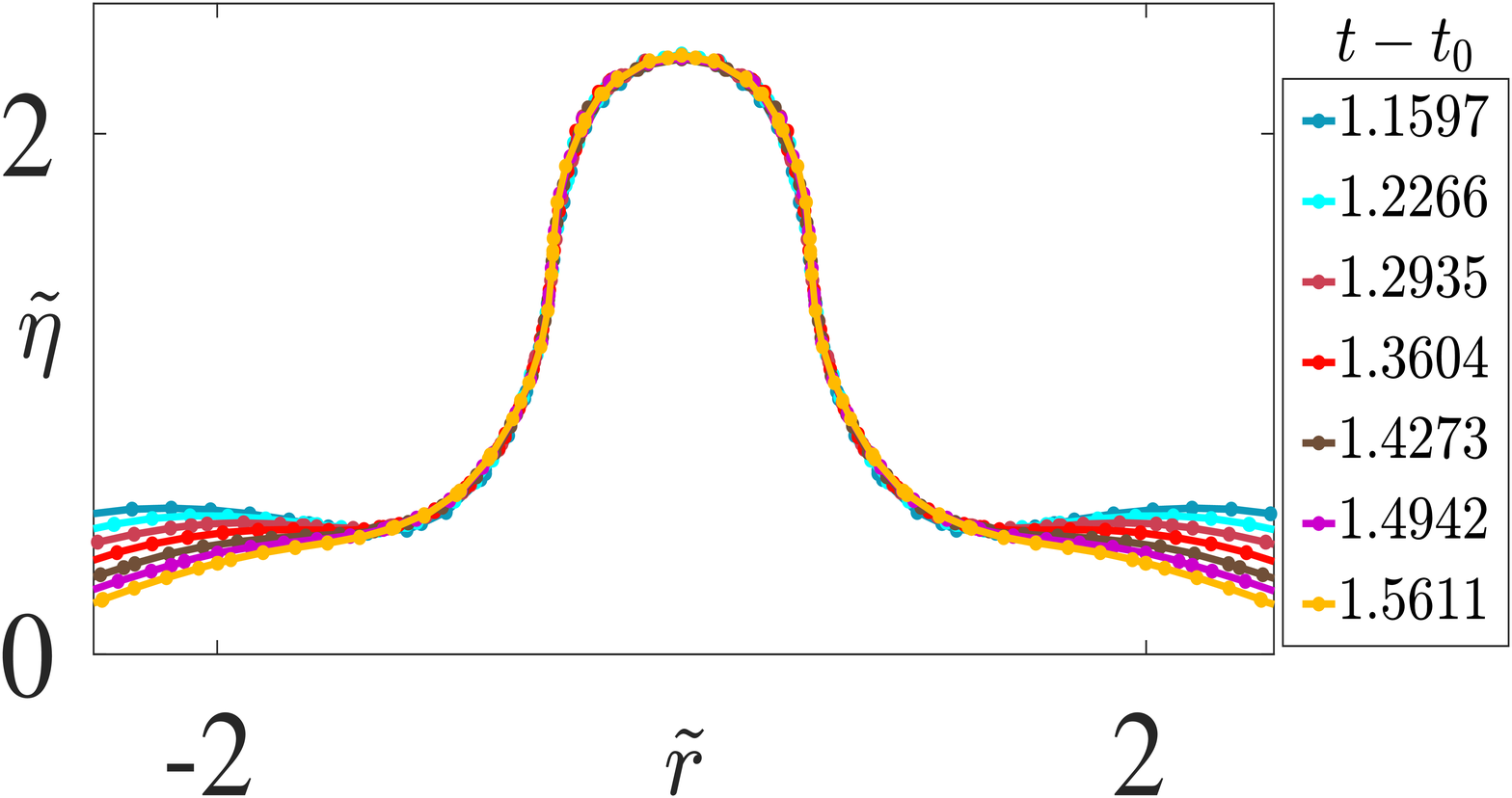}\label{fig5d}}
	\caption{Left column : $\epsilon = 2.3, l_{15} = 47.9$, Right column : $\epsilon = 2.3, l_{24} = 74.18$, Upper rows: unscaled data, Lower rows: scaled data. The linear time period of the standing wave in units of $t$ is $2\pi$. The inset (top right panel) shows the instant of dimple formation $t_0$.}
	\label{fig5}
\end{figure}

As $\epsilon >> 1$, we approach the strongly nonlinear regime where the curvature at the symmetry axis, as a consequence of wave focussing, increases sharply (singular). This is reflected in the progressive shortening of radial extent of the dimple as $\epsilon$ is increased, see fig. \ref{fig4} panel b  (also see fig. \ref{fig2d}). In the context of jets generated from bubble bursting and Faraday waves, such local singularities have been studied and self-similar solutions obtained in their vicinity \citep{zeff2000singularity,duchemin2002jet,ganan2017revision,lai2018bubble}. We anticipate an analogous singularity in local curvature as $\epsilon$ is increased, generating sharply shooting jets. Resorting to dimensional analysis the functional dependence of $\hat{\eta}$ on the physical parameters of the problem is expressed as $\hat{\eta} = \hat{f}(\hat{r},\hat{t},\hat{a}_0,\hat{R}_0,\frac{T}{\rho},l_q)$. This may be re-written non-dimensionally as $\eta \equiv \frac{\hat{\eta}l_q}{\hat{R}_0} = f\left(\frac{l_q}{\hat{R}_0}\hat r,\sqrt{\frac{Tl_q^3}{\rho\hat{R}_0^3}}\hat t;\;\epsilon,l_q\right) = f(r,t;\epsilon,l_q)$. We hypothesize that for sufficiently large $\epsilon$ and $l_q$, the local curvature at the axis of symmetry becomes singular due to wave focussing. Consequently, the temporal evolution of the jet that emerges from the dimple must be independent of $\hat{R}_0$ and $l_q$ (locally in space). This can occur only if in a narrow region around the jet (where the local singularity is effectively felt), $\eta$ depends only on the combination $\frac{r^3}{t^2}$ i.e. independent of $l_q/\hat{R}_0$. We thus have the ansatz that $\tilde{\eta}(\tilde{r}) \equiv \frac{\eta - z_b}{\left(t - t_0\right)^{2/3}}= g\left(\tilde{r};\epsilon,l_q\right)$ \citep{gekle2010generation,lai2018bubble} where $\tilde{r} \equiv \frac{r}{\left(t - t_0\right)^{2/3}}$, $t_0$ is the instant of dimple formation and $z_b$ is the (scaled) location of the first minima nearest to the symmetry axis. The scaled data is depicted in fig. \ref{fig5} for $\epsilon=2.3$ and $l_{15}$ and $l_{24}$. The upper row depicts the unscaled data at different times while the lower shows the scaled data. The excellent collapse in a region around the jet is apparent.
 
The similarity scales above are the \cite{keller1983surface} scales noted in context of the singularity at pinch-off for a fluid wedge. Fig. \ref{fig5} shows that while the temporal evolution of the jet happens self-similarly, initial and boundary parameters are not forgotten as the scaled interface shape $\tilde{\eta} = f(\tilde{r})$, depends on the parameters $(\epsilon,l_q)$. This is analogous to \cite{keller1983surface} where the self-similar wedge shape is not universal, but depends on the initial wedge angle. 
\begin{figure}
	\centering
	\includegraphics[scale=0.1]{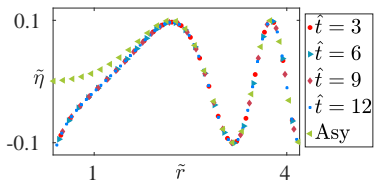}
	\caption{$\hat{\eta}(\hat{r},0) =\hat{V}_0\frac{\delta(\hat{r})}{2\pi\hat{r}}$. The interface evolves self-similarly at all time. (Asy) Asymptotic solution from stationary phase for large $\hat{r}$ and $\hat{t}$.}
	\label{fig6}
\end{figure}
Interestingly, the localised self-similar behaviour observed here in the strongly nonlinear regime may also be found in the linearised regime \citep{keller1983surface}. For a radially unbounded domain (i.e. no confining walls) and with only an initial interface deformation (zero surface impulse), the solution to the axisymmetric, linearised Cauchy-Poisson problem \citep{debnath1994nonlinear,kang2019gravity} is $\hat{\eta}(\hat{r},\hat{t})=\int_{0}^{\infty}k\mathrm{J}_0(k\hat{r})\bar{\eta}_0(k)\cos (\omega\hat{t})dk$, $\bar{\eta}_0(k)$ being Hankel transform of $\hat{\eta}(\hat{r},0)$. This solution lacks adequate length-scales and unless the initial condition introduces these, the resultant waves evolve self-similarly. A classic example is for pure gravity waves in two dimensions. This solution (due to Cauchy \& Poisson) is discussed by \citet{lamb1924hydrodynamics}, art. 238. Applying  \cite{lamb1924hydrodynamics} and \cite{debnath1994nonlinear} arguments to pure capillary waves on a radial pool, we considered a localised initial perturbation viz. $\hat{\eta}(\hat{r},0)=\frac{\hat{V}_0}{2\pi \hat{r}}\delta(\hat{r})$, where $\hat{V}_0^{1/3}$ is the only length scale in this initial condition. Dimensional reasoning predicts that instead of $\hat{\eta}$ being a function of $\hat{r}$ and $\hat{t}$ separately, it should be possible to express the scaled interface in terms of a single scaled variable i.e. $\frac{\hat{\eta} T^{'} \hat{t}^2}{\hat{r}\hat{V}_0} = \psi\left(\frac{\hat{r}}{\left(T^{'}\right)^{1/3}\hat{t}^{2/3}}\right)$ where $T^{'} \equiv \frac{T}{\rho}$. The functional form of $\psi(\cdot)$ may be obtained by solving the aforementioned integral numerically. Alternatively, the asymptotic form for $\psi(\cdot)$ is deduced from the stationary phase technique ($\hat{r},\hat{t}\rightarrow\infty$ with $\frac{\hat{r}}{\hat{t}}$ fixed) \citep{debnath1994nonlinear} (supplementary material). This asymptotic solution predicts
$\tilde{\tilde{\eta}} \equiv \frac{\hat{\eta}T^{'}\hat{t}^2}{\hat{V}_0\hat{r}}\sim \frac{2\sqrt{2}}{9\pi} \sin\left(\frac{4}{27} \tilde{\tilde{r}}^{3}\right)$ with $\tilde{\tilde{r}}\equiv \frac{\hat{r}}{\left(T'\right)^{1/3}\hat{t}^{2/3}}$
and we note the Keller-Miksis scale in this. Fig. \ref{fig6} compares the stationary phase solution (Asy) with the numerical solution to this problem at different time instances, showing self-similar behaviour. Addition of further length scales into the initial condition breaks this self-similarity. Consider linear capillary waves generated from a volume conserving perturbation \citep{miles1968cauchy,debnath1994nonlinear}, $\hat{\eta}(\hat{r},0)=\hat{d}\left(1-\frac{\hat{r}^2}{\hat{a}^2}\right)\exp\left(-\frac{\hat{r}^2}{\hat{a}^2}\right)$. This initial condition has two length scales viz. $\hat{d}$ and $\hat{a}$ and dimensional analysis predicts three groups viz. $\pi_1\equiv$ $\frac{\hat{\eta}(T')^3\hat{t}^6}{\hat{d}\hat{a}^4\hat{r}^5}$, $\pi_2\equiv$ $\frac{\hat{r}}{\left(T'\right)^{1/3}\hat{t}^{2/3}}$ and $\pi_3\equiv$ $\frac{\hat{a}^2\hat{r}^4}{(T')^2\hat{t}^4}$, implying from stationary phase, $\pi_1\sim \frac{8\sqrt{2}}{729}\exp\left(-\frac{4}{81}\pi_3\right)\sin\left(\frac{4}{27}\pi_2^3\right)$
which is not a self-similar result (supplementary material).

In conclusion, we have employed a single Bessel mode to produce dimples in the weakly nonlinear regime and jets in the strongly nonlinear regime. The choice of initial condition is crucial, as it allows us to theoretically explain wave focussing and radial inward motion of the capillary humps in terms of (nonlinear) generation of new eigenmodes. Our first principles calculation establishes that it is not necessary to invoke viscosity or gravity to explain a dimple or jet, although viscosity will retain a strong effect on the jet velocity. The inviscid self-similarity (first kind) observed in our simulations retains memory of boundary and initial parameters in contrast to the one observed in capillary pinch-off of an axisymmetric drop, where memory of initial parameters are lost \citep{PhysRevLett.80.704}. \\

\textbf{Acknowledgements:} Financial support from DST-SERB grants CRG/2020/003707 \& EMR/2016/000830 and IRCC (IIT Bombay) are gratefully acknowledged. LK acknowledges support from the Prime Minister's Research Fellowship (PMRF), Govt. of India. The tenure of SB at IIT Bombay, supported by PMRF is acknowledged.

\section*{Declaration of interests}
The authors report no conflict of interest.
\bibliographystyle{jfm}%
\bibliography{jfm}


\end{document}